\documentclass[12pt]{article}
\usepackage{amssymb,amsmath}
\usepackage{graphicx}
\usepackage{epsfig}
\usepackage{epstopdf}
\usepackage{latexsym}
\usepackage{psfrag}
\usepackage{booktabs}
\usepackage{bbm}

\usepackage[toc]{appendix}

\usepackage{color}
\usepackage{datetime}
\usepackage[
      colorlinks=false,
      linkcolor=darkblue,  
      urlcolor=blue,    
      filecolor=blue,     
      citecolor=red,
linktocpage=true,
      pdfstartview=FitV,
      bookmarksopen=true    
      ]{hyperref}

\ifpdf
\fi

\numberwithin{equation}{section}

\def\coeff#1#2{\relax{\textstyle {#1 \over #2}}\displaystyle}
\def\ds{\displaystyle}

\def\IC{\mathbb{C}}
\def\IH{\mathbb{H}}
\def\IP{\mathbb{P}}
\def\IR{\mathbb{R}}
\def\ZZ{\mathbb{Z}}

\def\cB{{\cal B}}

\def\cF{{\cal F}}

\def\cL{{\cal L}}

\def\cN{{\cal N}}

\def\Neql#1{{\cal N}\!=\!{#1}}

\def\nBPS#1{$\frac{1}{#1}$-BPS}

\newcommand{\dd}{d}				

\DeclareMathOperator{\sgn}{\mathrm{sgn}}	

\definecolor{cardinal}{rgb}{0.6,0,0}
\definecolor{darkgreen}{rgb}{0,0.5,0}
\definecolor{golden}{rgb}{0.92, 0.7, 0}
\definecolor{midnight}{rgb}{0, 0, 0.5}
\definecolor{darkblue}{rgb}{0.2, 0, 0.8}

\begin{document}  

\begin{titlepage}
 
\bigskip
\bigskip
\bigskip
\bigskip
\begin{center} 
{\Large \bf   New Supersymmetric Bubbles on $AdS_3\times S^3$}

\bigskip
\bigskip 

{\bf Nikolay Bobev${}^{(1)}$, 
Benjamin E. Niehoff${}^{(2)}$ and Nicholas P. Warner${}^{(2)}$ \\ }
\bigskip
${}^{(1)}$
Simons Center for Geometry and Physics\\
Stony Brook University\\
Stony Brook, NY 11794, USA\\
\vskip 5mm
${}^{(2)}$
Department of Physics and Astronomy \\
University of Southern California \\
Los Angeles, CA 90089, USA  \\
\bigskip
nbobev@scgp.stonybrook.edu,~bniehoff@usc.edu,~warner@usc.edu  \\
\end{center}

\bigskip
\bigskip

\begin{abstract}

\noindent  We find a large new class of explicit solutions preserving four supercharges in six-dimensional supergravity. The solutions are determined by solving a linear system of equations on a four-dimensional K\"ahler base studied by LeBrun. For particular choices of the parameters, we find regular backgrounds that are asymptotic to the near-horizon limit of the D1-D5-P black string. Holography implies that these backgrounds should be dual to \nBPS{4}  states in the D1-D5-P CFT.

\end{abstract}

\end{titlepage}


\tableofcontents

\section{Introduction}

Quite a number of important new results have come out of the systematic investigation of general BPS solutions with the same asymptotic structure at infinity as a given black hole or black ring.  While the primary goal of this program has, of course, been to evolve a deeper understanding of the structure of black-hole microstates, this work has also resulted in several  discoveries that have proven to have an even broader significance.   First, there was the realization that the BPS equations for five-dimensional, $\cN=2$ supergravity coupled to vector multiplets  are, in fact, linear \cite{Bena:2004de}.   This not only led to a dramatic extension of the known families of solutions, primarily through the superposition of multiple components,  but also led to the discovery and systematic construction of bubbled microstate geometries \cite{Bena:2005va, Berglund:2005vb, Bena:2007kg}.  The last few months have also seen a further surprising development in that it has now been shown that the corresponding BPS equations in six dimensions can also be reduced essentially to a linear system \cite{Bena:2011dd}. This observation is only now just beginning to be exploited and it is one of the purposes of this paper to use the results in \cite{Bena:2011dd} to obtain some new six-dimensional BPS solutions.

A second spin-off of the investigation of microstate geometries has been the whole non-BPS, or almost-BPS, program (see, for example, \cite{Gimon:2007mh,Goldstein:2008fq,Bena:2009ev,Bellucci:2009qv,Bena:2009en,Bena:2009qv,Bena:2009fi,Bobev:2009kn,Bobev:2011kk, Vasilakis:2011ki,Dall'Agata:2011nh}). The idea here has been to show that there are also {\it linear} systems of equations that govern families of non-supersymmetric solutions.  These solutions tend to be characterized as the superposition of fluxes, geometry and D-branes for which subsystems are supersymmetric but the supersymmetries of the  subsystems are incompatible with one another, rendering the entire solution non-supersymmetric.  One of the beauties of this approach is that the supersymmetry breaking is very controlled and is determined by the separation of the subsystems whose supersymmetries are incompatible (see, for example, \cite{Vasilakis:2011ki}).  The linearity of the underlying equations means that it is, at least in principle, straightforward (though often technically arduous) to assemble large families of examples of such non-BPS geometries. 

One class of such non-BPS solutions involves assembling families of mutually BPS branes that would be supersymmetric in flat space but then putting them in a curved background whose holonomy breaks the supersymmetry. To solve the equations of motion the background must be chosen in a very specific manner and can be either based upon a Ricci-flat or electrovac solution in four Euclidean dimensions.   The five-dimensional background is then obtained by adding a time coordinate, angular momenta, warp factors and more electromagnetic fields.  Each of these additions is determined through a linear system of equations.

New families of such five-dimensional supergravity solutions were investigated in \cite{Bena:2009fi, Bobev:2009kn}, where the starting point was a four-dimensional  Euclidean electrovac background on which the supergravity equations of motion were rendered solvable. A particularly interesting class of such four-dimensional bases that are K\"ahler and have vanishing Ricci scalar were utilized in this context in \cite{Bobev:2011kk}. As with many other examples, it simplifies the problem greatly if one assumes that the K\"ahler manifold has at least a $U(1)$ isometry and the most general local form of such metrics was determined in \cite{LeBrun:1991}. These ``LeBrun metrics''  are defined by two functions, one of which must satisfy the Affine Toda equation and the other of which must essentially be harmonic in a background defined by the Affine Toda solution.   This generalizes  the story of how the local form of a hyper-K\"ahler metric with a $U(1)$ isometry is determined by a single function that solves the Affine Toda equation  \cite{Boyer:1982mm,DasGegenberg}.  If one makes a simple choice for the solution to the Affine Toda equation then one  is led to the ``LeBrun-Burns metrics,''  which provide a simple, explicit class of K\"ahler metrics on $\IC^2$ with $n$ points blown up.  These metrics are structurally similar to the Gibbons-Hawking (GH) metrics \cite{Gibbons:1979zt} (which are hyper-K\"ahler) except that the $\IR^3$ sections and the harmonic functions on these sections are now replaced by the hyperbolic space, $\mathbb{H}^3$, and its harmonic functions.   The LeBrun-Burns metrics are also asymptotic to $\IR^4$ but the special $U(1)$ isometry of the manifold does {\it not} act in a manner that matches the  tri-holomorphic $U(1)$ action on a Gibbons-Hawking metric that is similarly  asymptotic to $\IR^4$.    Thus the solutions obtained from the LeBrun-Burns metrics will be intrinsically different from those with a GH base.  

The electromagnetic field for the LeBrun metrics has components that involve the K\"ahler form.  This means that the electromagnetic field does not vanish at infinity and, as was shown in  \cite{Bobev:2011kk}, the corresponding supergravity solutions are not asymptotically flat. Indeed, the natural, physical boundary conditions for such solutions correspond to the $AdS \times S$ geometries of the near-horizon regions of a black hole or black ring.  Several such solutions were found in  \cite{Bobev:2011kk} but regularity in five space-time dimensions imposed requirements  that significantly restricted  the solutions.  In retrospect, it is now evident that the natural asymptotic structure that was emerging from the analysis of the solutions based upon the LeBrun-Burns metrics was that they wanted to be asymptotic to $AdS_3 \times S^3$ in six dimensions.  As we will discuss in this paper, the naturally emerging fall-off for the warp factors directly leads to  $AdS_3 \times S^3$  when the five-dimensional solutions of M-theory on $T^6$ are recast as six-dimensional solutions of IIB supergravity on $T^4$.  

Another reason for revisiting the LeBrun metrics from the six-dimensional perspective is the similarity of some of the system of non-BPS equations in five-dimensions \cite{Bena:2009en, Bena:2009fi, Bobev:2011kk}  and the BPS equations in six dimensions \cite{Gutowski:2003rg,Cariglia:2004kk, Bena:2011dd}.   We will show here that even though the solutions based upon the LeBrun metrics in five dimensions are not supersymmetric, the solutions {\it are supersymmetric} in the six-dimensional, IIB duality frame.  More generally, the LeBrun solutions are non-supersymmetric in  M-theory and are only supersymmetric in the particular IIB frame in which the electromagnetic field of the LeBrun base is used to give the momentum charge to the overall solution.  The reason for this is that the surviving supersymmetry, or the frames that define them,  necessarily have a charge under the $U(1)$ of the momentum charge fibration in six dimensions.  The supersymmetry is broken by the trivial KK reduction of the six-dimensional solution and then any trivial uplift of this solution, such as to M-theory, does not restore the supersymmetry. This is, perhaps, a little reminiscent of Scherk-Schwarz reduction on a circle \cite{Scherk:1978ta,Scherk:1979zr} but the latter explicitly introduces masses through  dependence of the fields on the extra dimensions whereas here the dependence on extra dimensions only arises in the supersymmetry and not in the fields themselves\footnote{See also \cite{Duff:1997qz} for a somewhat similar supersymmetry breaking mechanism.}.

It is important to note that the BPS conditions of six-dimensional supergravity lead to differential constraints on a four-dimensional Euclidean base which, so far, have only been solved on a hyper-K\"ahler manifold \cite{Gutowski:2003rg,Cariglia:2004kk,Bena:2011dd}. In fact, most explicit BPS solutions in six-dimensional supergravity constructed to date have a four-dimensional hyper-K\"ahler base with a tri-holomorphic $U(1)$.  That is, these solutions have been based upon  Gibbons-Hawking metrics\footnote{A notable exception is \cite{Bena:2007ju} where more general hyper-K\"ahler manifolds were considered.}. As we discuss in detail below, the LeBrun metrics represent a general class of K\"ahler but \emph{not} hyper-K\"ahler bases on which the six-dimensional BPS constraints can be solved and for  the LeBrun-Burns metrics these solutions are completely explicit.

In Section 2 we will review the non-BPS equations of motion resulting from the floating brane Ansatz of \cite{Bena:2009fi} and the non-BPS solutions with a LeBrun base studied in \cite{Bobev:2011kk}. In section 3 we show how to recast these solutions as supersymmetric backgrounds in six dimensions and show that the six-dimensional BPS equations exactly reduce to the non-BPS equations of motion in five dimensions. Section 4 contains a discussion of the explicit new six-dimensional solutions on a LeBrun-Burns base and a short summary of their asymptotics and the conditions for regularity. In Section 5 we conclude with a summary and some questions for future study.
  
\section{The non-BPS solutions based upon the LeBrun metrics}
\label{nBPSsect}

\subsection{The non-BPS equations}
\label{nBPSsystem}

We will work with $\Neql 2$, five-dimensional ungauged supergravity coupled to two vector multiplets. This theory can also be viewed as a consistent truncation of eleven-dimensional supergravity on $T^6$.  Our conventions will be those of \cite{Bobev:2011kk}. The metric of the five-dimensional supergravity solution has the form:  
\begin{equation}
ds_5^2 ~=~ -Z^{-2} \,(dt + k)^2 ~+~ Z \, ds_4^2  \,,
\label{metAnsatz}
\end{equation}
where the base metric, $ds_4^2$, will ultimately be taken to be the LeBrun K\"ahler electrovac background.    The three background Maxwell fields are given by the  vector potentials:
\begin{equation}
A^{(I)}   ~=~  - Z_I^{-1}\, (dt +k) + B^{(I)}  \,,
\label{AAnsatz}
\end{equation}
where $B^{(I)}$ is a one-form on the base $ds_4^2$ and $I=1,2,3$.   One can introduce the magnetic two-form field strengths:
\begin{equation}
	\Theta^{(I)} ~\equiv~ d B^{(I)}  \label{Bpots} \,.
\end{equation}

The ``floating brane'' Ansatz \cite{Bena:2009fi} fixes the two scalars in the vector multiplets in terms of the ratios $Z_I/Z_J$ and requires that we take the warp factor, $Z$,  to be given by:
\begin{equation}
Z ~\equiv~ \big( Z_1 \, Z_2 \, Z_3  \big)^{1/3}   \,.
\label{Zdefn}
\end{equation}

The four-dimensional base, $ds^2_4$, has to be an Euclidean electrovac solution 
\begin{equation}
{R}_{\mu\nu} = \coeff{1}{2}\, \big( \cF_{\mu\rho} {\cF_{\nu}}^{\rho} -  \coeff{1}{4}\, g_{\mu\nu} \cF_{\rho\sigma} \cF^{\rho\sigma} \big)\,,
 \label{electrovacequation}
\end{equation}
where all quantities are computed in the four-dimensional base metric. The Maxwell field, $\mathcal{F}$, can be decomposed as:
\begin{equation} 
\cF = \Theta^{(3)} - \omega^{(3)}_{-}\,,
\label{Fdecomp}
\end{equation}
where $\Theta^{(3)}$ is self-dual and $\omega^{(3)}_{-}$ is anti-self-dual.  The Maxwell equations $d \cF = d * \cF =0$ imply that $\Theta^{(3)}$ and $\omega^{(3)}_{-}$ are harmonic.  As the notation implies, the decomposition \eqref{Fdecomp} defines the magnetic two-form field strength $\Theta^{(3)}$ in \eqref{Bpots}.

The supergravity equations of motion can be written as a linear system \cite{Bena:2009fi}:   
\begin{eqnarray}
 \hat \nabla^2 Z_1 &=&   *_4  \big[ \Theta^{(2)} \wedge \Theta^{(3)}   \big]    \,,  \qquad \big(\Theta^{(2)} -    *_4 \Theta^{(2)} \big)  ~=~  2  \, Z_1 \, \omega^{(3)}_- \,,  \label{ZMax1} \\
\hat \nabla^2 Z_2&=& *_4  \big[ \Theta^{(1)} \wedge \Theta^{(3)}   \big]  \,,  \qquad
 \big(\Theta^{(1)} -   *_4 \Theta^{(1)} \big)    ~=~ 2 \, Z_2 \, \omega^{(3)}_-  \,, \label{ZMax2}
\end{eqnarray}
and
\begin{eqnarray}
 \hat \nabla^2 Z_3  &=&  *_4  \big[ \Theta^{(1)} \wedge \Theta^{(2)}   ~-~    \omega^{(3)}_- \wedge(  dk   -  *_4  dk  )  \big]    \,, 
\label{Z3eqn} \\
 dk    ~+~   *_4 dk    &=&  \frac{1}{2} \, \sum_I \, Z_I \,  \big(\Theta^{(I)} +  *_4 \Theta^{(I)} \big)     \,,  \label{keqn} 
\end{eqnarray}
where $*_4$ and $\hat \nabla^2$ are the Hodge operator and the Laplacian on $ds^2_4$. The choice of the electrovac solution defines the base metric, $ds_4^2$, and one uses \eqref{Fdecomp} to read off $\Theta^{(3)}$ and $\omega_-^{(3)}$.  Equations  \eqref{ZMax1} and \eqref{ZMax2} are thus two linear coupled equations for $Z_1$ and $\Theta^{(2)}$ and $Z_2$ and $\Theta^{(1)}$ respectively.  Once these equations are solved, the angular momentum one-form $k$ and the metric function $Z_3$ are obtained as solutions to the system of linear equations \eqref{Z3eqn} and \eqref{keqn}.

\subsection{The  solutions based upon the LeBrun metrics}
\label{LeBrunMet}

The LeBrun metric, \cite{LeBrun:1991}, is the most general K\"ahler metric with a $U(1)$ isometry and a vanishing Ricci scalar. It takes the form
\begin{equation}
ds^2_{4}~=~  w^{-1}\, (d\tau + A)^2 ~+~  w\, (e^{u} (dx^2+dy^2) +dz^2) \,,
\label{LBmet}
\end{equation}
where $u$ and $w$ are two functions of $(x,y,z)$ which obey the $su(\infty)$ Toda equation and its linearized form:
\begin{eqnarray}
&&\partial_{x}^2 \,u ~+~ \partial_{y}^2 \,u ~+~ \partial_z^2\, (e^u) ~=~  0 \,,  \label{Toda} \\
&&\partial_{x}^2 \,w ~+~ \partial_{y}^2 \,w ~+~ \partial_z^2\,(e^u\, w) ~=~ 0\,.\label{linearToda}
\end{eqnarray}
The one-form, $A$, satisfies:  
\begin{equation}
dA ~=~ \partial_{x}w ~dy\wedge dz ~-~  \partial_{y}w~ dx\wedge dz ~+~ \partial_z(e^u w) ~dx \wedge dy\,, \label{Adefn}
\end{equation}
and the integrability of this differential, $d^2 A=0$, is equivalent to the equation (\ref{linearToda}). The  K\"ahler form is:
\begin{equation}
J ~=~ (d\tau + A) \wedge dz ~-~   w\, e^u \,  dx\wedge dy  \,. \label{Kform}
\end{equation}

It is convenient to introduce frames:
\begin{equation}
 \hat e^0 ~\equiv~  w^{-{1 \over 2}} \,  (d\tau + A)   \,,  \qquad  \hat  e^1 ~\equiv~  w^{ {1 \over 2}} \,  e^{ {u \over 2}} \, dx   \,,   
\qquad \hat  e^2 ~\equiv~   w^{ {1 \over 2}} \,  e^{ {u \over 2}} \, dy  \,,  \qquad \hat  e^3 ~\equiv~  w^{ {1 \over 2}} \,  dz  \,,  \label{frames}
\end{equation}
and the self-dual forms
\begin{eqnarray}
&& \Omega^{(1)}_+  ~\equiv~  e^{- {u \over 2}}  ( \hat e^0 \wedge \hat e^1 ~+~\hat  e^2 \wedge \hat e^3) ~=~ (d\tau + A) \wedge dx ~+~   w\,   dy \wedge dz \,,  \nonumber   \\
&&  \Omega^{(2)}_+  ~\equiv~   e^{- {u \over 2}}  ( \hat e^0 \wedge \hat e^2 ~-~ \hat e^1 \wedge \hat e^3) ~=~ (d\tau + A) \wedge dy ~-~   w\,   dx \wedge dz \,,    \label{sdforms}\\
&&  \Omega^{(3)}_+  ~\equiv~   ( \hat e^0 \wedge \hat e^3 ~+~ \hat e^1 \wedge \hat e^2) ~=~ (d\tau + A) \wedge dz ~+~   w\,  e^u\, dx \wedge dy \,.  \nonumber 
\end{eqnarray}
We will also frequently denote the coordinates by $\vec y \equiv (y_1, y_2, y_3) =(x,y,z) $.  With our choice of duality convention, the K\"ahler form, $J$, in  (\ref{Kform}) is anti-self-dual.

As discussed in \cite{Bobev:2011kk} the LeBrun metrics are four-dimensional electrovac solutions with a Maxwell field, $\cF$, given by (\ref{Fdecomp}):
\begin{equation}
\Theta^{(3)}  ~=~    \frac{1}{ 2 } \, \sum_{a=1}^3 \bigg( \partial_a \bigg( {\partial_z u \over w} \bigg)\bigg) \,  \Omega^{(a)}_+ \,, \qquad\qquad  \omega^{(3)}_- ~=~ J \,.  \label{Max3}
\end{equation}
The two-form, $\Theta^{(3)} $,  has a vector potential given by:
\begin{equation}
B^{(3)}  ~=~    \frac{1}{2} \,    \bigg[   - \bigg(  {\partial_z u \over w} \bigg)\, (d \tau + A)  ~+~ (\partial_y u) \, dx  ~-~ (\partial_x u)  \, dy  \, \bigg]  \,.   \label{Bvec3}
\end{equation}

The differential equations for the five-dimensional, non-BPS solutions based upon the LeBrun metrics were extensively reduced in  \cite{Bobev:2011kk}.  It was first shown that the three five-dimensional gauge fields are determined by:
\begin{equation}
\Theta^{(1)}~=~ Z_2 \, J ~+~   \sum_{a=1}^3 p^{(1)}_a \,   \Omega^{(a)}_+ \,, \qquad
\Theta^{(2)}~=~ Z_1 \, J ~+~   \sum_{a=1}^3 p^{(2)}_a \,   \Omega^{(a)}_+   \,, \label{Thetaforms}
\end{equation}
\begin{equation}
Z_1 ~=~ \frac{1}{2}\,   \Big(  {K^{(2)} \, \partial_z u \over w} \Big) ~+~L_1\,, \qquad Z_2 ~=~  \frac{1}{2}\,   \Big( { K^{(1)} \, \partial_z u \over w} \Big) ~+~L_2   \,, \label{Z12form}
\end{equation}
where 
\begin{eqnarray}
&& p^{(1)}_1 ~=~  \partial_x \Big( {K^{(1)} \over w} \Big)  \,, \qquad  p^{(1)}_2 ~=~  \partial_y \Big( {K^{(1)} \over w} \Big)  \,, \qquad p^{(1)}_3 ~=~   -\, Z_2  ~+~ \partial_z \Big( {K^{(1)} \over w} \Big)   \,, \label{pfns1}  \\
&&p^{(2)}_1 ~=~  \partial_x \Big( {K^{(2)} \over w} \Big)  \,, \qquad  p^{(2)}_2 ~=~  \partial_y \Big( {K^{(2)} \over w} \Big)  \,, \qquad p^{(2)}_3 ~=~   -\, Z_1  ~+~ \partial_z \Big( {K^{(2)} \over w} \Big)   \,.  \label{pfns2} 
\end{eqnarray}
The functions $L_1$ and $L_2$ are only required to be solutions of (\ref{linearToda}), that is:
\begin{equation}
\partial_{x}^2 \,L_I ~+~ \partial_{y}^2 \,L_I ~+~ \partial_z^2\,(e^u\, L_I ) ~=~ 0\,,   \qquad I =1,2 \,,\label{Leqns1}
\end{equation}
and, given these solutions, the functions $K^{(1)}$ and $K^{(2)}$ are determined by the linear equations:
\begin{eqnarray}
&&\partial_{x}^2 \, K^{(1)}  ~+~ \partial_{y}^2 \,K^{(1)}  ~+~ \partial_z \,(e^u\, \partial_z \,K^{(1)} ) ~=~  2\,  \partial_z \,(e^u\, w\,   L_2 )   \,, \label{lin2a}  \\
&&\partial_{x}^2 \,K^{(2)}  ~+~ \partial_{y}^2 \,K^{(2)}  ~+~ \partial_z \,(e^u\, \partial_z \,K^{(2)} ) ~=~  2\,  \partial_z \,(e^u\, w\,   L_1 )   \,. \label{lin2b} 
\end{eqnarray}

The last part of the solution then has the form:
\begin{equation}
k ~\equiv~ \mu \, (d \tau + A)  ~+~    \omega \,,  \qquad Z_3 ~=~    {K^{(1)} \,  K^{(2)}  \over w}  ~+~L_3 \,,   \label{kZ3forms}
\end{equation}
where $\omega = \vec\omega \cdot d \vec y$ and
\begin{equation}
\mu ~=~   -\frac{1}{2}\, \Big(  {K^{(1)} \,  K^{(2)} \, \partial_z u  \over w^2} \Big) - \frac{1}{2}\, \Big(  {K^{(1)} \,  L_1  + K^{(2)} \,  L_2   \over w} \Big) ~-~ \frac{1}{4}\, \Big(  {\partial_z u \, L_3  \over w} \Big) ~+~ M \,.  \label{muform}
\end{equation}
The functions $L_3$ and $M$ must satisfy the following linear equations:
\begin{eqnarray}
&&\partial_{x}^2 \, M + \partial_{y}^2 \,  M  + \partial_z \,(e^u\, \partial_z \, M ) =   \partial_z \,(e^u \,  L_1 \, L_2 )   \,,  \label{lin3a}  \\
&& \partial_{x}^2 \, L_3  +  \partial_{y}^2 \, L_3  +  e^u\, \partial_z^2 \, L_3 =    -2 \,  e^u \big[  2\, w \,(- L_1 L_2 + \partial_z M) + L_1 \, \partial_z K^{(1)} + L_2 \, \partial_z K^{(2)}  \big]  \,,  \label{lin3b} 
\end{eqnarray}
and the components of $\vec \omega$ are determined from:
\begin{eqnarray}
(\partial_y \, \omega_z -  \partial_z \, \omega_y ) ~+~ (M \partial_x w - w \partial_x M)  &+&\frac{1}{2} \, \sum_{I=1}^2 (K^{(I)} \partial_x L_I -  L_I  \partial_x K^{(I)}) \notag\\ 
 &+& \frac{1}{4} \, \big((\partial_z u) \,  \partial_x L_3 - L_3 \partial_x (\partial_z u)  \big) ~=~ 0 \,,   \label{omx}  \\
- (\partial_x \, \omega_z -  \partial_z \, \omega_x ) ~+~ (M \partial_y w - w \partial_y M)  &+&\frac{1}{2} \, \sum_{I=1}^2 (K^{(I)} \partial_y L_I -  L_I  \partial_y K^{(I)}) \notag\\ 
 &+& \frac{1}{4} \, \big((\partial_z u) \,  \partial_y L_3 - L_3 \partial_y (\partial_z u)  \big) ~=~ 0  \,,  \label{omy} 
 \end{eqnarray}
 \begin{eqnarray}
 (\partial_x \, \omega_y -  \partial_y \, \omega_x ) &+& (M \partial_z (e^u \, w) - e^u \, w  \,  \partial_z M)  ~+~ \frac{1}{2} \, \sum_{I=1}^2 (K^{(I)} \partial_z (e^u \, L_I)  -   e^u \, L_I   \partial_z K^{(I)}) \notag\\ 
 &+& \frac{1}{4} \, \big((\partial_z e^u) \,  \partial_z L_3 - L_3 \partial_z^2 (e^u)  \big)~+~ 2\, e^u \, w\, L_1 \,L_2 ~=~ 0  \,.  \label{omz} 
\end{eqnarray}
The integrability of the equations for $\vec \omega$ is implied by the differential equations satisfied by all the other background functions.

\section{The six-dimensional solutions}

\subsection{The BPS equations in six dimensions}
\label{SixDBPS}

The six-dimensional system we study is $\Neql 1$ minimal supergravity coupled to one anti-self-dual tensor multiplet  and this may be viewed as  arising from a consistent truncation of IIB supergravity on $T^4$.  Upon trivial dimensional reduction, the six-dimensional theory gives rise to precisely the theory used in Section \ref{nBPSsect}: $\Neql 2$, five-dimensional  supergravity coupled to two vector multiplets.  In the six-dimensional theory, the graviton multiplet contains a self-dual tensor field and so the entire bosonic sector consists of the graviton, the dilaton and an unconstrained $2$-form gauge field with a $3$-form field strength\footnote{Our analysis could be extended to include solutions of IIB supergravity on $K3$ and thus to theories with more tensor multiplets.  In five dimensions this would correspond to $\Neql 2$ theories with more vector multiplets.  It should be straightforward to generalize our results to such systems and we leave a detailed discussion on this for future work.}.

Supersymmetric solutions of this supergravity theory necessarily have a very constrained form of the metric \cite{Gutowski:2003rg}: 
\begin{equation}
ds^2 =  -2 H^{-1} (dv+\beta) \big(du + \hat\omega ~+~ \coeff{1}{2}\, \widehat\cF\, (dv+\beta)\big) ~+~  H \, ds_4^2(\cB)\,.   \label{sixmet1}
\end{equation}
where the  metric on the four-dimensional base,  $\cB$, is written in terms of components as:
\begin{equation}
 ds_4^2~=~  h_{mn} dx^m dx^n \,. 
 \end{equation}
As was shown in \cite{Gutowski:2003rg} the functions which determine the six-dimensional background are independent of $u$,  that is, $\partial_{u}$ is an isometry.   One should note that we are using slightly different conventions from \cite{Gutowski:2003rg} in the metric signature of (\ref{sixmet1}) and in the definition of the Hodge dual.  We adopt the more standard convention:
\begin{equation} 
*_{n}(e^{i_1}\wedge\ldots \wedge e^{i_p}) = \frac{1}{(n-p)!}\epsilon^{i_1\ldots i_p}\,_{j_1\ldots j_{n-p}}e^{j_1}\wedge\ldots \wedge e^{j_{n-p}}  \,. \label{sixDdual}
\end{equation} 

As in five dimensions, the six-dimensional BPS solution can be encoded in a reduced set of fields:  three functions, denoted  by  $\widehat Z_1$, $\widehat Z_2$ and $\widehat\cF$;  three two forms,   $\widehat \Theta^{(I)}$, and an angular momentum one-form, $\hat\omega$, all defined on the base, $\cB$. Details of how these fields  encode the six-dimensional fields   can be found in \cite{Bena:2011dd}.  Here we will work purely with these reduced fields except that the functions and fields  in  \cite{Bena:2011dd}  will now been given hats, { }$\widehat { }${   }, so as to avoid confusion with the non-BPS objects in the foregoing section.     

The  essential difference between the five-dimensional supergravity and the six-dimensional one is that one of the Maxwell fields of the five-dimensional theory has been promoted to a Kaluza-Klein field while the other two Maxwell fields encode the self-dual and anti-self-dual parts of the $3$-form field strength.  Indeed, as we will show, the third Maxwell field in five dimensions, encoded by $Z_3$ and $B^{(3)}$, is elevated to the metric function, $\cF$, and the one-form, $\beta$, in the six-dimensional metric (\ref{sixmet1}).   The warp factor, $H$, and the dilaton, $\hat{\phi}$, are related to the $\widehat Z_I$:
\begin{equation}
 H  ~\equiv~  \sqrt{ \widehat Z_1   \widehat Z_2} \,, \qquad\qquad e^{2\sqrt{2} \,\hat{\phi}} ~\equiv~ \ds\frac{\widehat{Z}_1}{\widehat{Z}_2}  \,.  \label{Hdefn}
\end{equation}

The six-dimensional BPS conditions can be reduced to differential equations on the base, $\cB$, and, to this end, we introduce the restricted exterior derivative, ${\tilde{d}}$, acting on a $p$-form,   $\Phi \in \Lambda^p ({\cal B})$, by:
\begin{eqnarray} 
 \Phi &=& {1 \over p!} \, \Phi_{m_1 \dots m_p} (x,v) \, dx^{m_1} \wedge \ldots \wedge  dx^{m_p} \,, \\
 {\tilde{d}} \Phi &\equiv&  {1 \over (p+1)!} \, (p+1) \, {\partial \over \partial x^{[q}} \Phi_{m_1 \dots m_p]} \,  dx^q \wedge  dx^{m_1}
 \wedge \ldots \wedge dx^{m_p}\,.
\end{eqnarray} 
and we define a Kaluza-Klein covariant differential operator, $D$,  by:
\begin{equation} 
D \Phi ~\equiv~ {\tilde{d}} \Phi ~-~ \beta \wedge {\dot{\Phi}},  \label{betaFS}
\end{equation} 
where we denote a derivative with respect to $v$ by a dot.  The field strength, $ \widehat\Theta^{(3)}  \equiv  D \beta $,   is then required to satisfy the self-duality condition: 
\begin{equation} 
 \widehat\Theta^{(3)} ~\equiv~    *_4  \widehat\Theta^{(3)}  \,. \label{betacond}
\end{equation} 

The supersymmetry conditions imply that the base  is ``almost hyper-K\"ahler''  in that there are three anti-self-dual $2$-forms, 
\begin{equation}
 J^{(A)}  ~\equiv~ \coeff{1}{2}\,   {J^{(A)}}_{mn}  \, dx^m  \wedge dx^ n \,,
 \end{equation}
that satisfy the quaternionic algebra: 
\begin{equation}
{J^{(A)}}{}^m{}_p {J^{(B)}}{}^p{}_n = \epsilon^{ABC}\,{J^{(C)}}{}^m{}_n ~-~  \delta^ {AB} \,\delta^m_n \,.
\label{Jalg} 
 \end{equation}
These forms are also required to satisfy the differential identity:
\begin{equation}
\tilde d J^{(A)}  ~=~    \partial_v \big (\beta \wedge J^{(A)})  \label{Jcond} \,,
 \end{equation}
where $\partial_v\Phi$ denotes the Lie derivative of a quantity $\Phi$
with respect to the tangent vector $\partial \over \partial v$.  

Given this $v$-dependent structure, one can define the anti-self-dual $2$-forms, $\psi$ and $\hat \psi$, by:
\begin{equation}
\psi  ~\equiv~ H \, \hat \psi  ~\equiv~  \coeff{1}{16} \, H\,  \epsilon^{ABC}  \, J^{(A)}{}^{mn} \dot {J}^{(B)}{}_{mn} \, J^{(C)} \label{psidefn}  \,,
\end{equation}
This form measures the failure of self-duality of the  the $\widehat \Theta^{(a)}$, $a=1,2$:
\begin{equation}
*_4 \widehat\Theta^{(1)}     ~=~ \widehat\Theta^{(1)}  -  2 \,  \widehat Z_2\, \hat \psi  \,,    \qquad *_4 \widehat\Theta^{(2)}     ~=~ \widehat\Theta^{(2)}  -  2 \,  \widehat Z_1\, \hat \psi  \,.  \label{Thetaduals}
\end{equation}
In particular, the anti-self-dual parts of the $\widehat\Theta^{(a)}$ are proportional to $\hat \psi$. 
 
Note that if one makes the identifications $\widehat\Theta^{(1)}=\widehat \Theta^{(2)}$  and $\widehat Z_1=\widehat Z_2$ then the three-form flux is self-dual in six dimensions and the dilation vanishes.  This reduces the theory reduces to minimal six-dimensional supergravity.

With these definitions, the following equations determine  $\widehat Z_a$ and $\widehat \Theta^{(a)}$:
\begin{equation}
\tilde d \widehat\Theta^{(2)} ~=~  \partial_v  \big[ {-}\coeff{1}{2} *_4 (D \widehat Z_1 + \dot{\beta} \widehat Z_1)~+~  \beta \wedge \widehat\Theta^{(2)} \big] \,, \qquad 
D *_4 (D\widehat Z_1 + \dot{\beta} \widehat Z_1) =  2 \,\widehat\Theta^{(2)} \wedge  D \beta\,, \label{Z1eqn}
\end{equation}
\begin{equation}
\tilde d \widehat\Theta^{(1)}~=~  \partial_v  \big[ {-}\coeff{1}{2} *_4 (D \widehat Z_2 + \dot{\beta} \widehat Z_2)~+~  \beta\wedge \widehat\Theta^{(1)} \big] \,, \qquad
D *_4 (D \widehat Z_2 + \dot{\beta} \widehat Z_2) =  2 \,\widehat\Theta^{(1)} \wedge  D \beta\,. \label{Z2eqn}
\end{equation}

It is convenient to write the final system of equations in terms of a new one-form, $L$, defined by: 
\begin{equation}
L ~\equiv~  \dot{\hat{\omega}}~+~ \coeff{1}{2}\, \dot \beta\widehat\cF  ~-~  \coeff{1}{2}\,  D \widehat\cF\ \,.  \label{Lphidefn}
\end{equation}
The function $\widehat\cF$ and the angular momentum vector, $\hat\omega$, are then determined by:
\begin{eqnarray}
{-}*_4 D *_4 L &=& \coeff{1}{2}\, H h^{mn}\partial_v^2 (H h_{mn}) + \coeff{1}{4}\, \partial_v (H h^{mn}) \,\partial_v (H h_{mn}) - 2\, \dot{\beta}_m\, L^m + 2\, H^2 \,\dot{\hat\phi}^2  \nonumber\\
&& - 2*_4 \Big[\, \widehat\Theta_1 \wedge \widehat\Theta_2 ~-~  H^{-1}  \psi  \wedge D \hat\omega \, \Big] \,,  \label{Feqn} \\
D \hat\omega + *_4 D \hat\omega  &=&    2\,Z_1\, \widehat\Theta_1 +  2\,Z_2\, \widehat\Theta_2  ~-~  \widehat\cF \, D\beta - 4\, H\, \psi \nonumber \\
&=&    2\,Z_1\, \big (\widehat\Theta_1 - Z_2\, \hat \psi\big) +  2\,Z_2\, \big(\widehat\Theta_2 - Z_1\, \hat \psi\big) ~-~  \widehat\cF \, D\beta      \label{angmomeqn} \,,
\end{eqnarray}
where  the dilaton, $\hat{\phi}$, is defined in \eqref{Hdefn}.

The analysis of the supersymmetries  requires a choice of frames and it is simplest if one uses the null system:
\begin{equation} 
e^+ \equiv H^{-1} \big(dv + \beta \big)  \,,  \qquad
e^- ~\equiv~  du +\hat\omega   + \coeff{1}{2} \,\widehat\cF H \, e^+  \,, \qquad 
e^a = H^{1 \over 2} \tilde{e}^a{}_m dx^m \,,  \label{eqn:frames}
\end{equation} 
in which the metric may be written
\begin{equation} 
ds^2 ~=~   -2 e^+ e^- ~+~ \delta_{ab}\, e^a  \, e^b \,.
\end{equation} 
To further pin down the choice of frames on the base one chooses frames in which the forms defining the almost hyper-K\"ahler structure have constant coefficients.  To be specific, if one lowers the indices using the metric, $h_{mn}$, on the base then, one choses frames on the base, $\tilde{e}^a$, in (\ref{eqn:frames}) so that 
\begin{equation}
  J^{(1)}   ~\equiv~  \tilde e^0 \wedge \tilde  e^1 ~-~ \tilde e^2 \wedge \tilde e^3 \,,  \qquad   J^{(2)} ~\equiv~ \tilde e^0 \wedge \tilde e^2 ~+~ \tilde e^1 \wedge \tilde e^3  \,, \qquad  J^{(3)}  ~\equiv~  \tilde e^0 \wedge \tilde e^3 ~-~ \tilde e^1 \wedge \tilde  e^2    \,.    \label{hyperK2}\\ 
\end{equation}
If the fields satisfy the  BPS equations  then, with the foregoing choice of frames, the supersymmetries are, in fact, constants  \cite{Gutowski:2003rg,Cariglia:2004kk}:
\begin{equation} 
\partial_\mu \epsilon ~=~ 0  \,. \label{consteps}
\end{equation} 
%

\subsection{The Lebrun metrics as a base for BPS solutions}
\label{ConvertLeBrun}

Here we show that the LeBrun metrics can be used as a four-dimensional base for constructing six-dimensional BPS solutions of the form described above. 

The first step is to find a self-dual Maxwell field on the base and for this we choose 
\begin{equation}
\widehat\Theta^{(3)}=D\beta  ~=~ \coeff{1}{2} \,  \Theta^{(3)} \,, \qquad
\beta  =   \coeff{1}{2} \, B^{(3)}  ~=~    \frac{1}{4} \,    \bigg[   - \bigg(  {\partial_z u \over w} \bigg)\, (d\tau + A)  ~+~ (\partial_y u) \, dx  ~-~ (\partial_x u)  \, dy  \, \bigg]  \,.   \label{betachoice}
\end{equation}
There is an obvious anti-self-dual almost-hyper-K\"ahler structure on the LeBrun base:
\begin{eqnarray}
&& \widehat J^{(1)}   ~\equiv~  \hat e^0 \wedge \hat  e^1 ~-~ \hat e^2 \wedge \hat e^3 ~=~  e^{ {u \over 2}}  ((d\tau + A) \wedge dx ~-~   w\,   dy \wedge dz ) \,,  \nonumber   \\
&&  \widehat J^{(2)} ~\equiv~ \hat e^0 \wedge \hat e^2 ~+~ \hat e^1 \wedge \hat e^3  ~=~ e^{ {u \over 2}}  ((d\tau + A) \wedge dy ~+~   w\,   dx \wedge dz) \,,    \label{hyperK1}\\
&&  J^{(3)}  ~\equiv~ J  ~=~  \hat e^0 \wedge \hat e^3 ~-~ \hat e^1 \wedge \hat  e^2  ~=~ (d\tau + A) \wedge dz ~-~   w\,  e^u\, dx \wedge dy \,,  \nonumber 
\end{eqnarray}
where the frames are defined in (\ref{frames}) and $J$ is the original K\"ahler form.   However $\widehat J^{(1)} , \widehat J^{(2)}$ and $J^{(3)}$  are $v$-independent and  only $J^{(3)}$ is closed and so they do not satisfy the differential constraint (\ref{Jcond}).  On the other hand, if one defines a rotating form of these structures: 
\begin{equation}
J^{(1)}   ~\equiv~   \cos (2\, v)\,  \widehat J^{(1)} ~-~ \sin (2\, v) \,\widehat J^{(2)} \,, \qquad   J^{(2)}   ~\equiv~   \sin (2\, v) \, \widehat J^{(1)} ~+~ \cos (2\, v) \, \widehat J^{(2)}  \,,   \label{rotJ}
\end{equation}
one finds that the $J^{(A)}$ are a set of almost hyper-K\"ahler structures that do indeed obey  (\ref{Jcond}).  The fact that this elementary modification works is a very special property of the LeBrun family of metrics and does not work in other familiar examples of four-dimensional metrics, like the Israel-Wilson metrics  used as a base for five or six-dimensional supergravity solutions \cite{Bena:2009fi}.

With this choice for the  $J^{(A)}$, it is easy to verify that   
\begin{equation}
 \hat \psi  ~\equiv~  \coeff{1}{16} \,   \epsilon^{ABC}  \, J^{(A)}{}^{mn} \dot {J}^{(B)}{}_{mn} \, J^{(C)} ~=~  J^{(3)}   ~=~  J  \label{psires}  \,.
\end{equation}
From this and  (\ref{Max3})   one immediately sees that the duality conditions, (\ref{Thetaduals}), of the BPS system are precisely the same as (\ref{ZMax1}) and (\ref{ZMax2}), which are the non-BPS duality  conditions.  This suggest the obvious identifications:
\begin{equation}
\widehat  Z_a    ~=~   Z_a    \,,  \qquad \widehat \Theta^{(a)}     ~=~  \Theta^{(a)}   \,, \qquad a =1,2\,,   \label{ZTidents}  
\end{equation}
where all of these functions and forms will be taken to be $v$-independent. Equations (\ref{Z1eqn}) and (\ref{Z2eqn}) then imply that $\widehat \Theta^{(a)}$ is closed, which is consistent with the non-BPS conditions (\ref{Bpots}).  With the identifications \eqref{betachoice} and \eqref{ZTidents}, the equations in   (\ref{Z1eqn}) and (\ref{Z2eqn})  reduce to the other non-BPS equations in (\ref{ZMax1}) and (\ref{ZMax2}).
 
Finally, (\ref{Feqn}) and (\ref{angmomeqn}) reduce to (\ref{Z3eqn}) and (\ref{keqn}) if one makes the identifications
\begin{equation}
\widehat\cF   ~=~   -4 \, Z_3    \,,  \qquad   \hat\omega   ~=~  2\, k\,.    \label{Z3kidents}  
\end{equation}
One can then rewrite the metric  (\ref{sixmet1})  as a standard fibration of the $v$-circle over a five dimensional space-time and upon reduction on this $v$-circle one obtains precisely the metric (\ref{metAnsatz})  provided one sets  $u= 2t$. 

Thus the non-BPS ``floating brane" solutions in five dimensions based upon the LeBrun metrics found in \cite{Bobev:2011kk} can be recast as supersymmetric solutions in the six-dimensional framework.  This appears to contradict the belief that the non-BPS systems do not have supersymmetry.  However it is relatively easy to resolve this apparent inconsistency.  

One should note that the constancy of the Killing spinors (\ref{consteps}) was contingent upon being in a system of frames in which the almost hyper-K\"ahler forms have constant coefficients (\ref{hyperK2}).  However, the differential constraints on the $J^{(A)}$ required that we pass to the system of rotating structures, (\ref{rotJ}) and so the frames, $\tilde e^a$, for the six-dimensional constant spinors must be related to the standard, $v$-independent frames, $\hat e^a$, of the LeBrun base via:
\begin{equation}
\tilde e^1   ~=~   \cos (2\, v)\,  \hat e^1 ~-~ \sin (2\, v) \,  \hat e^2  \,, \qquad  \tilde e^2  ~=~    \cos (2\, v)\,  \hat e^2 ~+~ \sin (2\, v) \,  \hat e^1  \,,   \label{rotfram}
\end{equation}
One could, of course, work in six dimensions with the frames, $\hat e^a$,  and transform everything using the foregoing frame rotation.   One would then find that the  supersymmetries {\it necessarily depend upon $v$}.  It is for this reason that {\it trivial dimensional reduction} to five dimensions {\it breaks the supersymmetry}.

More generally, if one works purely in five dimensions, or in any setting, like M-theory, where there is no non-trivial Kaluza-Klein fibration,  then there is no way to preserve the supersymmetry because the  fiber dependence that is essential to the supersymmetry cannot be realized. Thus it is only in the six-dimensional theory and its IIB uplift that the solutions with a LeBrun base can be rendered supersymmetric.

\section{Explicit solutions}

The system of differential equations in Section \ref{LeBrunMet} can be explicitly solved for a large class of   LeBrun-Burns spaces and such solutions were analyzed in great detail in \cite{Bobev:2011kk}. However, the focus there was primarily upon finding solutions that were regular in five-dimensions and this imposed very stringent boundary conditions that greatly reduced the possibilities. As we showed in the previous section, solving the differential equations in Section \ref{LeBrunMet} leads to explicit BPS solutions in six dimensions. In view of this we will revisit the results of \cite{Bobev:2011kk} and show that there is a rich new class of BPS solutions of the six-dimensional theory.

\subsection{The LeBrun-Burns metrics}
\label{LBBsols}

The general solution of the $su(\infty)$-Toda equation that determines the function, $u$, is extremely difficult to find due to the non-linear nature of the equation.  However, there is a very interesting class of backgrounds, the LeBrun-Burns metrics, that arise  from a simple  solution to \eqref{Toda}:
\begin{equation}
u  ~=~   \log (2 \, z)\,.   \label{LBBuform}
\end{equation}
To study this class of spaces  it is convenient to define 
\begin{equation}
 z  ~\equiv~   \coeff{1}{2}\, \zeta^2 \,, \qquad V ~\equiv~ e^u \, w ~=~  2\,  z\, w    ~=~   \zeta^2 \, w \,.   \label{zVform}
\end{equation}
The LeBrun-Burns metric can then be written as
\begin{equation}
ds^2_4  ~=~  \zeta^2 \Big[  V^{-1} \, (d\tau+A)^2 + V \Big(  \frac{dx^2 + dy^2 + d \zeta ^2}{\zeta ^2} \Big) \Big]  \label{fourmet}\,.
\end{equation}
The three-dimensional metric is the standard constant-curvature metric on the hyperbolic plane, $\IH^3$: 
\begin{equation}
ds^2_{\IH^3} ~=~  \ds\frac{dx^2 + dy^2 + d\zeta^2}{\zeta ^2} \,. \label{HypMet}
\end{equation}
The equations (\ref{linearToda}) and  (\ref{Adefn}) that define the four-dimensional base imply that $V$ is a harmonic function on the hyperbolic plane and that $A$ is an appropriate one-form on $\IH^3$: 
\begin{equation}
\nabla^2_{\IH^3} V ~=~  0\,,  \qquad\qquad dA ~=~ *_{\IH^3} dV \,. \label{Hypconds}
\end{equation}

The most explicitly-known solutions are  axi-symmetric and it is therefore convenient to introduce polar coordinates:
\begin{equation}
x ~=~\rho \sin \theta \cos \phi\,, \qquad y ~=~ \rho \sin \theta \sin \phi\,,   \qquad      \zeta ~=~ \rho \cos \theta  \,.  \label{polars}
\end{equation}
and define the functions
\begin{equation}
H_{i} ~\equiv~   \frac{1}{\sqrt{(\rho^2+c_i^2)^2-4 \zeta^2 c_i^2}}\,,  \qquad G_{i} ~\equiv~    (\rho^2 +c_i^2) \, H_{i} ~-~ 1 \,, \qquad
D_{i} ~\equiv~   (\rho^2 -c_i^2)  \, H_{i}  \,,
\end{equation}
for some parameters, $c_i \ne 0$.   One can then solve (\ref{linearToda}) and  (\ref{Leqns1}) by taking
\begin{equation}
V~=~ \varepsilon_0  ~+~ \sum_{j=1}^N \, q_j \,  G_j \,,  \qquad L_{a} ~=~ \frac{1}{\zeta^2} \big(\ell_a^{0} +\ds\sum_{i=1}^{N}\ell_{a}^{i}G_{i}\big)~, \qquad a=1,2\,,  \label{VLform}
\end{equation}
for some free parameters $ \varepsilon_0$, $q_j $,  $\ell_a^{0}$ and $\ell_{a}^{i}$.

As shown in \cite{Bobev:2011kk}, the rest of the solution is then given by:

\begin{eqnarray}
K^{(1)} &=& k_1^{0} + \frac{\beta_1}{\rho^2}+\ds\sum_{i=1}^{N}k_{1}^{i}H_{i} - V L_{2}+ 4\rho^2 \ds\sum_{i,j=1}^{N} q_{i} \ell_2^{j}H_iH_j \,, \label{K1res}\\
K^{(2)} &=& k_2^{0} + \frac{\beta_2}{\rho^2}+\ds\sum_{i=1}^{N}k_{2}^{i}H_{i} -  V L_{1} + 4\rho^2 \ds\sum_{i,j=1}^{N} q_{i} \ell_1^{j}H_iH_j \,, \label{K2res}\\
M &=& m_{0}+ \ds\frac{\gamma}{\rho^2} +\ds\sum_{i=1}^{N}m_{i}H_{i} -  \ds\frac{\zeta^2}{2} L_{1}L_2 + 2\rho^2 \ds\sum_{i,j=1}^{N} \ell_1^{i} \ell_2^{j}H_iH_j \,,\label{Mres}
\end{eqnarray}
\begin{eqnarray} 
\label{L3resi}
&& L_3 = \ell_3^{0} + \ds\sum_{i=1}^{N} \ell_3^i G_i - \zeta^2 VL_1 L_2 + \ds\sum_{i=1}^{N} (2(\varepsilon_0 - Q)m_{i} +(\ell_1^0-\Lambda_1) k_1^i+(\ell_2^0-\Lambda_2) k_2^i)H_{i} \notag\\
&&+ \beta_3\ds\frac{\zeta^2}{\rho^4}+(2(\varepsilon_0 - Q)\gamma +(\ell_1^0-\Lambda_1) \beta_1+(\ell_2^0-\Lambda_2)\beta_2)\ds\frac{1}{\rho^2}+ 2\gamma \ds\sum_{i=1}^{N} \ds\frac{q_{i} }{c_i^2} \ds\frac{\rho^{-2}-H_i}{H_i}~, \notag\\
&&+ \ds\sum_{i=1}^{N} (2q_{i}m_{i} + \ell_1^i k_1^i + \ell_2^i k_2^i) (\eta^2-\zeta^2+c_i^2)H_{i}^2 + \ds\sum_{i\neq j =1}^{N} \ds\frac{(2q_{i}m_{j} + \ell_1^i k_1^j + \ell_2^i k_2^j)}{c_i^2-c_j^2} \ds\frac{H_j-H_i}{H_i} \notag\\
&&+4\ds\sum_{i,j=1}^{N} ((\varepsilon_0 - Q) \ell_1^i \ell_2^j+ (\ell_1^0-\Lambda_1) q_i \ell_2^j + (\ell_2^0-\Lambda_2)q_i \ell_1^j)\rho^2 H_i H_j \\ 
&&+ 4 \ds\sum_{i,j,k=1}^{N} q_i \ell_1^j \ell_2^k \rho^2 (3\rho^2 - 4 \zeta^2+c_i^2+c_j^2+c_k^2) H_{i} H_j H_k \,, \notag
\end{eqnarray}
\begin{eqnarray} 
\omega  &=& \bigg[\,\omega_{0}+ \ds\frac{\beta_3}{2}  \frac{\sin^2\theta}{\rho^2} - \gamma\ds\sum_{i=1}^{N}  \frac{q_i}{c_i^2} D_i -  \sum_{j=1}^{N} \bigg( m_0q_j + k_1^0 \ell_1^j + k_2^0 \ell_2^j + \ds\frac{\ell_3^j}{2} \bigg)D_j\notag\\ 
&&-   \sum_{j=1}^{N} (2m_jq_j+k_1^j \ell_1^j + k_2^j \ell_2^j)\eta^2 H_j^2   -  \sum_{i\neq j =1}^{N} \ds\frac{(2q_{i} m_{j}+k_1^i \ell_1^j + k_2^i \ell_2^j)}{2(c_i^2-c_j^2)} (D_iD_j + 4\eta^2c_i^2 H_i H_j)  \notag \\
&& - 8 \ds\sum_{i,j,k=1}^{N} q_i \ell_1^j \ell_2^k \eta^2\rho^2 H_iH_jH_k \, \bigg]\, d \phi  \,,  \label{omegasoln}
\end{eqnarray}
where $\beta_J$, $\gamma$, $k_a^{0}$,  $k_a^i$, $m_0$, $m_i$,   $\ell_3^{0}$, $\ell_3^i$ and $\omega_0$ are also  free parameters and
\begin{equation}\label{QLL def}
Q ~\equiv~  \sum_{i=1}^{N} \, q_i\,, \qquad\qquad \Lambda_{1} ~\equiv~  \sum_{i=1}^{N}\, \ell_1^{i}\,,  \qquad\qquad \Lambda_{2} ~\equiv~    \sum_{i=1}^{N} \, \ell_2^{i}\,.
\end{equation}
Finally, the functions that appear in the metric and background fields can be expressed in terms of the functions above as:
\begin{eqnarray}
Z_1 &=& \ds\frac{K^{(2)}}{V} +L_1~, \qquad Z_2 ~=~ \ds\frac{K^{(1)}}{V} +L_2~, \qquad Z_3 ~=~ \ds\frac{\zeta^2 K^{(1)}K^{(2)}}{V} +L_3~, \\
\mu &=& M - \ds\frac{1}{2}\ds\frac{L_3}{V} - \ds\frac{1}{2} \ds\frac{\zeta^2(K^{(1)}L_1 + K^{(2)}L_2)}{V} - \ds\frac{\zeta^2 K^{(1)}K^{(2)}}{V^2}~.
\end{eqnarray}
%

\subsection{Asymptotics and Regularity}
\label{Regularity}

\subsubsection{Asymptotics at infinity}
\label{AsympInf}

To understand the asymptotic behavior of the general multi-center solution presented above we will study in some detail the spherically symmetric solution on a flat $\mathbb{R}^4$ base which corresponds to choosing $V=1$ for the function determining the Burns base. The sources for this solution lie at $(x, y, \zeta)=(0,0,0)$.  We will also set some of the electric potentials to zero: 
\begin{equation}
L_1 ~\equiv~L_2 ~\equiv~ 0\,. \label{BHsol1}
\end{equation}
The functions $K^{(I)}$ and $M$ are then homogeneous solutions to  $\cL_1 H =0$, where
\begin{equation}
\cL_1 H ~ \equiv~  \partial_{x}^2 H   ~+~ \partial_{y}^2 \, H  ~+~ \zeta^{-1} \, \partial_\zeta \, (\zeta  \partial_\zeta  H )   \,,  \label{diffop1} 
\end{equation}
and we take
\begin{equation}
  Z_1 ~=~  K^{(2)} ~=~   \frac{\beta_2}{  \rho^2}  \,, \qquad Z_2 ~=~   K^{(1)} ~=~      \frac{   \beta_1}{  \rho^2} \,,  \qquad M ~=~   \frac{\gamma}{  \rho^2}  \label{BHsol2}  \,, 
\end{equation}
where  $\beta_1, \beta_2 $ and $\gamma$ are constant parameters.  

It is easy to see that the rest of the functions in the solution are
\begin{eqnarray}
 Z_3 &=&  \ell_3^0 ~+~  \frac{2\, \gamma}{  \rho^2} ~+~ (\beta_1 \,\beta_2\,  +  \beta_3 )\,  \frac{ \cos^2 \theta}{ \rho^2}   \,,  \label{BHsol5} \\
 \mu &=&  - \frac{1}{2}\, (2\,\beta_1 \,\beta_2 +   \beta_3) \,   \frac{ \cos^2 \theta}{ \rho^2} \,, \qquad \omega ~=~ \frac{\beta_3}{2}   \, \frac{ \sin^2 \theta}{ \rho^2} \, d \phi\,.  \label{BHsol6}
\end{eqnarray}
The six-dimensional metric is then
\begin{multline}
ds^2 ~=~ - \ds\frac{\rho^2}{\sqrt{\beta_1\beta_2}} \, dv\left(2du -2(2\,\beta_1 \,\beta_2 +   \beta_3) \,   \frac{ \cos^2 \theta}{ \rho^2}d\tau+ 2\beta_3   \, \frac{ \sin^2 \theta}{ \rho^2} \, d \phi - 4Z_3 dv\right)\\
\qquad\qquad + \sqrt{\beta_1\beta_2} \,\ds\frac{d\rho^2}{\rho^2} +   \sqrt{\beta_1\beta_2} \,(d\theta^2 + \sin^2\theta d\phi^2 + \cos^2\theta d\tau^2)~.
\end{multline}
For generic values of the parameters above (in particular for $\ell_3^0\neq0$) this solution is asymptotic to a pp-wave type background at $\rho \to \infty$. However for $\ell_3^0=0$ and $\beta_3=-\beta_1\beta_2$ the metric becomes precisely the near horizon metric of a BPS D1-D5-P black string (see, for example, \cite{Giusto:2004id}). To have a precise identification of the parameters of our solution with the charges of the D1-D5-P string we performed a careful comparison with the 3-charge solutions in D1-D5-P frame discussed in \cite{Bena:2008dw}. We find the following identification 
\begin{equation}
Q_1=\beta_2~, \qquad Q_5=\beta_1~, \qquad Q_P=8\gamma~, \qquad J=\beta_1\beta_2~, \label{chargeident}
\end{equation}
where $Q_1$, $Q_5$ and $Q_P$ are D1, D5 and momentum charges of the black string and $J$ is its angular momentum. Note that the entropy of the black string is $S \sim\sqrt{Q_1Q_5Q_P-J^2}$ and we have the bound $Q_1Q_5Q_P\geq J^2$. It is also interesting to note that \eqref{chargeident} implies $J=Q_1Q_5$ which, for $Q_P=0$, is the condition for a maximally spinning  D1-D5  supertube \cite{Lunin:2002iz}. In general however we have $Q_P\neq 0$ and the condition $J=Q_1Q_5$ seems less natural. It will be very interesting to understand this relation between $J$, $Q_1$ and $Q_5$  from the point of view of the dual  D1-D5-P CFT.

\subsubsection{Asymptotics near the charge centers}
\label{RegCenters}

As one would expect, one can easily recover the solutions for multiple concentric black rings \cite{Bena:2004de,Elvang:2004ds,Gauntlett:2004qy} from our general multi-center solutions.  The details depend upon the behavior of the solution as $\rho_i \to 0$, where 
\begin{equation}
\rho_i ~\equiv~ \sqrt{ x^2 + y^2 + (\zeta-c_i)^2} \,. \label{rhoidefn}
\end{equation}
One can easily arrange that  all three $Z_I \sim \rho_i^{-1}$ and $V \sim  \rho_i^{-1} $ and then one finds that the metric opens up into a rotating  $AdS_3 \times S^3$ throat as $ \rho_i \to 0$ and each such center thus corresponds to a rotating black ring/string.  

Another possibility in six dimensions is that the geometry remains smooth as $ \rho_i \to 0$  in precisely the same manner that it does for a two-charge supertube \cite{Lunin:2002iz,Bena:2008dw}.  This requires three basic ingredients:  a)   $Z_1, Z_2 \sim \rho_i^{-1}$,  b)  $Z_3$ remains finite and  c)  the $v$-fiber combines with the $S^2$ in $\IH^3$ around $\rho_i =0$ so as to pinch off as an (orbifold of) $S^3$ as $\rho_i \to 0$.   

One can easily verify that the necessary conditions on the $Z$'s can be met.  For example, we can take
\begin{equation}
q_i = k_1^i = k_2^i = 0
\end{equation}
at some point $(0,0,c_i)$, but place no other restriction on the parameters of the solution.  Then
\begin{equation}
Z_1 \sim  \ds\frac{\ell_1^i}{c_i\rho_i} ~, \qquad\qquad Z_2 \sim  \ds\frac{\ell_2^i}{c_i\rho_i},
\end{equation}
and after some straightforward, yet tedious, algebra, we obtain
\begin{equation}
\begin{split}
Z_3 &\sim \frac{1}{\rho_i} \bigg( c_i \ell_3^i + (\epsilon_0 - Q) \frac{m_i}{c_i} - \frac{m_i}{c_i} \sum_{j \neq i} q_j \sgn (c_i^2 - c_j^2) \\
& \qquad \qquad + \frac{\ell_1^i}{c_i} \sum_{j \neq i} \Big[ (\epsilon_0 - Q) \ell_2^i + (\ell_2^0 - \Lambda_2) q_j \Big] \sgn (c_i^2 - c_j^2)  \\
& \qquad \qquad + \frac{\ell_2^i}{c_i} \sum_{j \neq i} \Big[ (\epsilon_0 - Q) \ell_1^i + (\ell_1^0 - \Lambda_1) q_j \Big] \sgn (c_i^2 - c_j^2)  \\
& \qquad \qquad - \frac{\ell_1^i}{c_i} \sum_{j,k \neq i} q_j \ell_2^k \sgn \big[ (c_i^2 - c_j^2) (c_i^2 - c_k^2) \big]  \\
& \qquad \qquad - \frac{\ell_2^i}{c_i} \sum_{j,k \neq i} q_j \ell_1^k \sgn \big[ (c_i^2 - c_j^2) (c_i^2 - c_k^2) \big] \bigg), 
\end{split}
\end{equation}
where the indices are summed over all other points $(0,0,c_j)$, and $Q, \Lambda_1, \Lambda_2$ are as in \eqref{QLL def}.  So we see that $Z_3$ can be made regular if the parameters are chosen to make this expression vanish.  To avoid problems with the metric signature, it is important that $\ell_1^i$ and $\ell_2^i$ have the same sign, which puts a further restriction on possible solutions.

However, solutions do exist, as this simple example shows:  Consider a two-center configuration, where at the first point
\begin{equation}
\ell_1^1 = \ell_2^1 = 0, \quad q_1 \neq 0,
\end{equation}
and at the second point
\begin{equation}
\ell_1^2 \equiv \ell_1, \quad \ell_2^2 \equiv \ell_2, \quad q_2 = \ell_3^2 = m_2 = 0.
\end{equation}
Also take
\begin{equation}
\ell_1^0 = a, \quad \ell_2^0 = b.
\end{equation}
Then at point 2, where $Z_1$ and $Z_2$ blow up, regularity of $Z_3$ requires
\begin{equation} \label{Z3 reg}
a \ell_1 + b \ell_2 - 2 \ell_1 \ell_2 = 0,
\end{equation}
which clearly has solutions.  More importantly, if $a$ and $b$ are both positive, then \eqref{Z3 reg} has solutions where $\ell_1$ and $\ell_2$ are either both positive or both negative.

However, the condition on the $v$-fibration cannot be satisfied.  Thus, while the metric can be made finite as $ \rho_i \to 0$, the metric is not regular because the surfaces with $ \rho_i \to \epsilon$, for $\epsilon \to 0$, have topology $S^2 \times S^1$ and not that of $S^3/\ZZ_p$.  To get a non-trivial fibration of the $v$-fiber requires the vector field, $\beta$, and its associated field strength $\widehat \Theta^{(3)}$ to have non-trivial flux through $2$-cycles in the base.  The fact that these fluxes are trivial arises from (\ref{Max3}) and the extremely simple, and non-singular choice we made for $u$ in (\ref{LBBuform}).  More general LeBrun metrics can certainly have such non-trivial fluxes, just as $\Theta^{(3)}$ can have non-trivial fluxes on GH bases, but the  structure  of LeBrun-Burns metrics precludes such supertubes.

Finally, we note that if one has $Z_1, Z_2, Z_3 \sim \rho_i^{-1}$ and $V$ finite then the solution is singular because  the metric defined by (\ref{sixmet1}) and (\ref{fourmet}) has the size of the $\tau$-circle diverging as $\rho_i \to 0$.  On the other hand, the metric is regular either when a) all the $Z_1, Z_2, Z_3$ are finite (as discussed in \cite{Bobev:2011kk}), or b) $Z_1, Z_2, V \sim \rho_i^{-1}$ and $Z_3$ is finite, in which case the metric opens up into a rotating $AdS_2 \times S^3 \times S^1$ throat.

\subsubsection{Ambipolar solutions}
\label{ambipolar}

We would like to point out that just as for BPS solutions of five-dimensional supergravity with a Gibbons-Hawking base \cite{Giusto:2004kj,Bena:2005va} we can use an ambipolar Burns base space  and obtain viable Lorentzian six-dimensional backgrounds. An ambipolar base is a four-dimensional base on which the signature changes signature from $(+,+,+,+)$ to $(-,-,-,-)$. On the Burns base this is achieved by having both positive and negative residues at the poles of the function $V$ \eqref{VLform}. Recall that the metric function $H$ is defined as
\begin{equation}
H = \sqrt{Z_1 \, Z_2} \,.
\end{equation}
The six-dimensional metric can be written
\begin{multline}
\dd s_6^2 = - \frac{1}{4 Z_3  \sqrt{Z_1 Z_2}} (\dd u + \omega)^2 + \frac{4 Z_3}{\sqrt{Z_1 Z_2} } \Big( \dd v + \beta - \frac{1}{4 Z_3} (\dd u + \omega) \Big)^2 \\ + \sqrt{Z_1 Z_2}  \,\zeta^2 \Big[  V^{-1} \, (d\tau+A)^2 + V \Big(  \frac{dx^2 + dy^2 + d \zeta ^2}{\zeta ^2} \Big) \Big]  \label{sixmet2}\,.
\end{multline}
Then we see that if a base is ambipolar the signature of the six-dimensional metric is left unchanged as long as $V, Z_1, Z_2, Z_3$ all change sign at the same locus. This precisely parallels the structure of the five-dimensional supergravity solutions with an ambipolar base \cite{Bena:2005va}.

\section{Conclusions}

We have studied a new class of BPS solutions of six-dimensional supergravity coupled to a tensor multiplet and these solutions can be trivially uplifted to supersymmeric solution of IIB supergravity on $T^4$. A key ingredient in our construction is a four-dimensional K\"ahler base with a $U(1)$ symmetry and vanishing Ricci scalar studied  by LeBrun. For a particular class of such four-dimensional metrics the BPS equations can be solved explicitly and one can find closed form expressions for the metric and the background fields. It is important to stress that these solutions provide the first examples of BPS backgrounds of six-dimensional supergravity that do not have a hyper-K\"ahler base. In fact, almost all explicit BPS solutions discussed previously have the very special Gibbons-Hawking base\footnote{To the best of our knowledge the only solutions with a more-general hyper-K\"ahler base are the ones in \cite{Bena:2007ju}.}.

The supersymmetry conditions of six-dimensional supergravity impose, amongst other things, a constraint, \eqref{Jcond}, on the four-dimensional base of the solution.  In contrast to the situation in five-dimensional supergravity, where this base has to be hyper-K\"ahler, it is not clear to us whether there is a simple geometric meaning of the more general constraint in six dimensions. It is quite conceivable that this constraint could be given a very interesting  meaning for some suitably arranged five-dimensional spatial geometry.   Our analysis clearly demonstrates that some K\"ahler manifolds can satisfy this constraint but we believe there will be a much more general class of geometries that can be used to construct six-dimensional BPS solutions.

For judicious choice of parameters our solutions are asymptotic, at infinity, to the near horizon geometry of the BPS D1-D5-P black string. It is certainly important to understand the microscopic brane configurations that source the solutions in more detail. Since the D1-D5-P black string geometry is asymptotically locally $AdS_3\times S^3$ one can apply holographic methods to uncover which states in the D1-D5-P CFT are dual to our regular solutions. The technology developed in \cite{Kanitscheider:2006zf} for the more restricted two-charge D1-D5 geometries will be certainly useful in this regard.   It will also be interesting to see if there is an efficient way to count our regular geometries by some generalization of the techniques used in \cite{Grant:2005qc, Rychkov:2005ji} to count two-charge sueprtubes or the \nBPS{2}  asymptotically $AdS_5\times S^5$ solutions of Lin-Lunin-Maldacena (LLM) \cite{Lin:2004nb}.

As we emphasized, the Killing spinors of our backgrounds will not survive a trivial dimensional reduction along the $v$-fiber and so supersymmetry will be broken in such a reduction. Moreover, a subsequent {\it trivial} uplift, like embedding the solution in M-theory will not restore the supersymmetry.  Since the six-dimensional solution is BPS, this means that five-dimensional non-BPS solutions are necessarily extremal because their mass is locked to their electric charges. Extremal non-BPS solutions in four and five dimensions have drawn a lot of attention recently and there is a large number of known multi-centered non-BPS solutions (see for example \cite{Bossard:2011kz}). It would be interesting to reduce our solutions to four dimensions and understand whether  the four-dimensional, axi-symmetric solutions    fit in one of the known classes of such solutions discussed in \cite{Bossard:2011kz} or whether the solutions discussed here provide a completely new system. Furthermore it will be interesting to explore the action of spectral flow \cite{Bena:2008wt} and more general U-duality symmetries of string theory on our solutions \cite{Dall'Agata:2010dy}.

Our solutions are not asymptotically flat and it would be nice to understand how to modify them such that we have a supergravity solution asymptotic to $\mathbb{R}^{1,5}$.   Although this is certainly an interesting question we expect that it will not be easy to answer it. For example, one does not know how to make the general \nBPS{2} LLM solutions in IIB asymptotically flat \cite{Lin:2004nb}.   On the other hand, there are certainly  more general solutions within reach that go beyond the ones constructed here.  As we remarked earlier, in (\ref{LBBuform}) we  made an extremely simple, non-singular choice for the solution, $u$,  of the Affine Toda equation and there are much richer possibilities.  Indeed, axi-symmetric solutions of the $su(\infty)$  Toda equation can be obtained by transforming solutions of the Laplace equation on $\mathbb{R}^3$ \cite{Ward:1990qt}.  It would be interesting to start from such solutions and see to what extent one can generate explicit BPS solutions. 

More generally, it would also be very interesting to address the question of classification of the asymptotically $AdS_3\times S^3$ solutions of six-dimensional supergravity which preserve four supercharges. This analysis was initiated in \cite{Martelli:2004xq, Liu:2004ru, Liu:2004hy} following the work of \cite{Lin:2004nb}. Such a classification may also lead to potential new insights as to how to count the regular \nBPS{4} solutions.  

The results we have presented here  not only  yield  insight into the relationship  between some families of BPS and almost-BPS, extremal solutions but also represent one of many possible new directions that can be explored from the perspective of six-dimensional supergravity.  It is evident that the linearity of the BPS equations in six dimensions \cite{Bena:2011dd} has opened up a rich new vein for research and will enable new, explicit constructions of families of BPS solutions.   

\bigskip
\bigskip
\leftline{\bf Acknowledgements}
\smallskip
We would like to thank Chris Beem, Iosif Bena, Frederik Denef, Sheer El-Showk, Stefano Giusto, Masaki Shigemori and Joan Sim\'on for useful conversations. NB and NPW are grateful to the Aspen Center for Physics where this work was initiated. NB appreciates the warm hospitality provided by the IPhT Saclay and the USC Department of Physics and Astronomy while part of this work was completed. The work of NB was supported in part by DOE grant DE-FG02-92ER-40697. The work of BN and NPW was supported in part by DOE grant DE-FG03-84ER-40168.  





\begin{thebibliography}{99}
   
\bibitem{Bena:2004de}
  I.~Bena and N.~P.~Warner,
 ``One ring to rule them all ... and in the darkness bind them?,''
  Adv.\ Theor.\ Math.\ Phys.\  {\bf 9}, 667 (2005)
  [arXiv:hep-th/0408106].
  
\bibitem{Bena:2005va}
  I.~Bena and N.~P.~Warner,
``Bubbling supertubes and foaming black holes,''
  Phys.\ Rev.\  D {\bf 74}, 066001 (2006)
  [arXiv:hep-th/0505166].
 
\bibitem{Berglund:2005vb}
  P.~Berglund, E.~G.~Gimon and T.~S.~Levi,
``Supergravity microstates for BPS black holes and black rings,''
  JHEP {\bf 0606}, 007 (2006)
  [arXiv:hep-th/0505167].

\bibitem{Bena:2007kg}
  I.~Bena and N.~P.~Warner,
``Black holes, black rings and their microstates,''
  Lect.\ Notes Phys.\  {\bf 755}, 1 (2008)
  [arXiv:hep-th/0701216].
 
\bibitem{Bena:2011dd}
  I.~Bena, S.~Giusto, M.~Shigemori and N.~P.~Warner,
``Supersymmetric Solutions in Six Dimensions: A Linear Structure,''
  arXiv:1110.2781 [hep-th].
 
\bibitem{Gimon:2007mh} 
  E.~G.~Gimon, F.~Larsen and J.~Simon,
  ``Black holes in Supergravity: The Non-BPS branch,''
  JHEP {\bf 0801}, 040 (2008)
  [arXiv:0710.4967 [hep-th]].
  
\bibitem{Goldstein:2008fq}
  K.~Goldstein and S.~Katmadas,
 ``Almost BPS black holes,''
  JHEP {\bf 0905}, 058 (2009)
  [arXiv:0812.4183 [hep-th]].

\bibitem{Bena:2009ev}
  I.~Bena, G.~Dall'Agata, S.~Giusto, C.~Ruef and N.~P.~Warner,
``Non-BPS Black Rings and Black Holes in Taub-NUT,''
  JHEP {\bf 0906}, 015 (2009)
  [arXiv:0902.4526 [hep-th]].
  
\bibitem{Bellucci:2009qv} 
  S.~Bellucci, S.~Ferrara, M.~Gunaydin and A.~Marrani,
  ``SAM Lectures on Extremal Black Holes in d=4 Extended Supergravity,''
  Springer Proc.\ Phys.\  {\bf 134}, 1 (2010)
  [arXiv:0905.3739 [hep-th]].

\bibitem{Bena:2009en}
  I.~Bena, S.~Giusto, C.~Ruef and N.~P.~Warner,
``Multi-Center non-BPS Black Holes - the Solution,''
  JHEP {\bf 0911}, 032 (2009)
  [arXiv:0908.2121 [hep-th]].
  
\bibitem{Bena:2009qv}
  I.~Bena, S.~Giusto, C.~Ruef and N.~P.~Warner,
``A (Running) Bolt for New Reasons,''
  JHEP {\bf 0911}, 089 (2009)
  [arXiv:0909.2559 [hep-th]].

\bibitem{Bena:2009fi}
  I.~Bena, S.~Giusto, C.~Ruef and N.~P.~Warner,
``Supergravity Solutions from Floating Branes,''
  JHEP {\bf 1003}, 047 (2010)
  [arXiv:0910.1860 [hep-th]].
    
\bibitem{Bobev:2009kn}
  N.~Bobev and C.~Ruef,
``The Nuts and Bolts of Einstein-Maxwell Solutions,''
  JHEP {\bf 1001}, 124 (2010)
  [arXiv:0912.0010 [hep-th]].
  
\bibitem{Bobev:2011kk}
  N.~Bobev, B.~Niehoff and N.~P.~Warner,
``Hair in the Back of a Throat: Non-Supersymmetric Multi-Center Solutions from K\"ahler Manifolds,''
  JHEP {\bf 1110}, 149 (2011)
  [arXiv:1103.0520 [hep-th]].

\bibitem{Vasilakis:2011ki}
  O.~Vasilakis and N.~P.~Warner,
``Mind the Gap: Supersymmetry Breaking in Scaling, Microstate Geometries,''
  JHEP {\bf 1110}, 006 (2011)
  [arXiv:1104.2641 [hep-th]].

\bibitem{Dall'Agata:2011nh} 
  G.~Dall'Agata,
  ``Black holes in supergravity: flow equations and duality,''
  arXiv:1106.2611 [hep-th].

\bibitem{LeBrun:1991}
C.~LeBrun ,
``Explicit Self-Dual Metrics on $\IC\IP_2 \# \ldots \IC\IP_2$,''
 J.\ Diff.\  Geom.  {\bf 34}  (1991) 223-253.
 %
 
\bibitem{Boyer:1982mm}
  C.~P.~Boyer and J.~D.~.~Finley,
  ``Killing Vectors In Self-dual, Euclidean Einstein Spaces,''
  J.\ Math.\ Phys.\  {\bf 23}, 1126 (1982).

%
\bibitem{DasGegenberg}
A. Das and J. Gegenberg, ``Stationary Riemannian  space-times with
self-dual curvature,'' Gen.\ Rel.\ Grav. {\bf 16}, (1984) 817. 

\bibitem{Gibbons:1979zt} 
  G.~W.~Gibbons and S.~W.~Hawking,
  ``Gravitational Multi - Instantons,''
  Phys.\ Lett.\ B {\bf 78}, 430 (1978).

\bibitem{Gutowski:2003rg}
  J.~B.~Gutowski, D.~Martelli and H.~S.~Reall,
``All supersymmetric solutions of minimal supergravity in six dimensions,''
  Class.\ Quant.\ Grav.\  {\bf 20}, 5049 (2003)
  [arXiv:hep-th/0306235].

\bibitem{Cariglia:2004kk}
  M.~Cariglia and O.~A.~P.~Mac Conamhna,
``The general form of supersymmetric solutions of N = (1,0) U(1) and  SU(2) gauged supergravities in six dimensions,''
  Class.\ Quant.\ Grav.\  {\bf 21}, 3171 (2004)
  [arXiv:hep-th/0402055].

\bibitem{Scherk:1978ta} 
  J.~Scherk and J.~H.~Schwarz,
  ``Spontaneous Breaking of Supersymmetry Through Dimensional Reduction,''
  Phys.\ Lett.\ B {\bf 82}, 60 (1979).
    
\bibitem{Scherk:1979zr} 
  J.~Scherk and J.~H.~Schwarz,
  ``How to Get Masses from Extra Dimensions,''
  Nucl.\ Phys.\ B {\bf 153}, 61 (1979).

\bibitem{Duff:1997qz} 
  M.~J.~Duff, H.~Lu and C.~N.~Pope,
  ``Supersymmetry without supersymmetry,''
  Phys.\ Lett.\ B {\bf 409}, 136 (1997)
  [hep-th/9704186].

\bibitem{Bena:2007ju} 
  I.~Bena, N.~Bobev and N.~P.~Warner,
  ``Bubbles on Manifolds with a U(1) Isometry,''
  JHEP {\bf 0708}, 004 (2007)
  [arXiv:0705.3641 [hep-th]].

\bibitem{Giusto:2004id} 
  S.~Giusto, S.~D.~Mathur and A.~Saxena,
  ``Dual geometries for a set of 3-charge microstates,''
  Nucl.\ Phys.\ B {\bf 701}, 357 (2004)
  [hep-th/0405017].

\bibitem{Bena:2008dw} 
  I.~Bena, N.~Bobev, C.~Ruef and N.~P.~Warner,
  ``Supertubes in Bubbling Backgrounds: Born-Infeld Meets Supergravity,''
  JHEP {\bf 0907}, 106 (2009)
  [arXiv:0812.2942 [hep-th]].

\bibitem{Lunin:2002iz} 
  O.~Lunin, J.~M.~Maldacena and L.~Maoz,
  ``Gravity solutions for the D1-D5 system with angular momentum,''
  hep-th/0212210.

\bibitem{Elvang:2004ds} 
  H.~Elvang, R.~Emparan, D.~Mateos and H.~S.~Reall,
  ``Supersymmetric black rings and three-charge supertubes,''
  Phys.\ Rev.\ D {\bf 71}, 024033 (2005)
  [hep-th/0408120].

\bibitem{Gauntlett:2004qy} 
  J.~P.~Gauntlett and J.~B.~Gutowski,
  ``General concentric black rings,''
  Phys.\ Rev.\ D {\bf 71}, 045002 (2005)
  [hep-th/0408122].

\bibitem{Giusto:2004kj} 
  S.~Giusto and S.~D.~Mathur,
  ``Geometry of D1-D5-P bound states,''
  Nucl.\ Phys.\ B {\bf 729}, 203 (2005)
  [hep-th/0409067].

\bibitem{Kanitscheider:2006zf} 
  I.~Kanitscheider, K.~Skenderis and M.~Taylor,
  ``Holographic anatomy of fuzzballs,''
  JHEP {\bf 0704}, 023 (2007)
  [hep-th/0611171].

\bibitem{Grant:2005qc} 
  L.~Grant, L.~Maoz, J.~Marsano, K.~Papadodimas and V.~S.~Rychkov,
  ``Minisuperspace quantization of 'Bubbling AdS' and free fermion droplets,''
  JHEP {\bf 0508}, 025 (2005)
  [hep-th/0505079].
   
\bibitem{Rychkov:2005ji} 
  V.~S.~Rychkov,
  ``D1-D5 black hole microstate counting from supergravity,''
  JHEP {\bf 0601}, 063 (2006)
  [hep-th/0512053].

\bibitem{Lin:2004nb} 
  H.~Lin, O.~Lunin and J.~M.~Maldacena,
  ``Bubbling AdS space and 1/2 BPS geometries,''
  JHEP {\bf 0410}, 025 (2004)
  [hep-th/0409174].

\bibitem{Bossard:2011kz} 
  G.~Bossard and C.~Ruef,
  ``Interacting non-BPS black holes,''
  Gen.\ Rel.\ Grav.\  {\bf 44}, 21 (2012)
  [arXiv:1106.5806 [hep-th]].

\bibitem{Bena:2008wt} 
  I.~Bena, N.~Bobev and N.~P.~Warner,
  ``Spectral Flow, and the Spectrum of Multi-Center Solutions,''
  Phys.\ Rev.\ D {\bf 77}, 125025 (2008)
  [arXiv:0803.1203 [hep-th]].
 
\bibitem{Dall'Agata:2010dy} 
  G.~Dall'Agata, S.~Giusto and C.~Ruef,
  ``U-duality and non-BPS solutions,''
  JHEP {\bf 1102}, 074 (2011)
  [arXiv:1012.4803 [hep-th]].

\bibitem{Ward:1990qt} 
  R.~S.~Ward,
  ``Einstein-Weyl spaces and SU(infinity) Toda fields,''
  Class.\ Quant.\ Grav.\  {\bf 7}, L95 (1990).

\bibitem{Martelli:2004xq} 
  D.~Martelli and J.~F.~Morales,
  ``Bubbling AdS(3),''
  JHEP {\bf 0502}, 048 (2005)
  [hep-th/0412136].

\bibitem{Liu:2004ru} 
  J.~T.~Liu, D.~Vaman and W.~Y.~Wen,
  ``Bubbling 1/4 BPS solutions in type IIB and supergravity reductions on S**n x S**n,''
  Nucl.\ Phys.\ B {\bf 739}, 285 (2006)
  [hep-th/0412043].
  
\bibitem{Liu:2004hy} 
  J.~T.~Liu and D.~Vaman,
  ``Bubbling 1/2 BPS solutions of minimal six-dimensional supergravity,''
  Phys.\ Lett.\ B {\bf 642}, 411 (2006)
  [hep-th/0412242].
   
\end{thebibliography}
\end{document}